\documentclass{article}

\usepackage{microtype}
\usepackage{graphicx}
\usepackage{subfigure}
\usepackage{booktabs} 

\usepackage{hyperref}

\usepackage[accepted]{icml2020}

\icmltitlerunning{Submission and Formatting Instructions for ICML 2020}

\usepackage{amsmath,amsthm,amssymb}
\usepackage{graphicx}
\usepackage{bm}
\usepackage{mathrsfs}
\usepackage{multirow}
\usepackage[all]{xy}
\usepackage{pbox}
\usepackage{verbatim}
\usepackage{bbold}
\usepackage{braket}
\usepackage{dsfont}
\usepackage{color,soul}
\usepackage{textcomp}

\newcommand{\Ss}{\mathcal{S}}
\newcommand{\As}{\mathcal{A}}
\newcommand{\Rs}{\mathcal{R}}

\newcommand{\dSs}{|\mathcal{S}|}
\newcommand{\dAs}{|\mathcal{A}|}

\newcommand{\vS}{V_{\mathcal{S}}}
\newcommand{\vR}{V_{\mathcal{R}}}

\usepackage{titling}

\icmltitlerunning{A Tensor Network Approach to Finite Markov Decision Processes}

\begin{document}

\twocolumn[
\icmltitle{A Tensor Network Approach to Finite Markov Decision Processes}




\begin{icmlauthorlist}
\icmlauthor{Edward Gillman}{nott,cmqs}
\icmlauthor{Dominic C. Rose}{nott,cmqs}
\icmlauthor{Juan P. Garrahan}{nott,cmqs}
\end{icmlauthorlist}

\icmlaffiliation{nott}{Department of Physics and Astronomy, University of Nottingham, Nottingham, NG7 2RD, UK}
\icmlaffiliation{cmqs}{Centre for the Mathematics and Theoretical Physics of Quantum Non-Equilibrium Systems, University of Nottingham, Nottingham, NG7 2RD, UK}

\icmlcorrespondingauthor{Edward Gillman}{edward.gillman@nottingham.ac.uk}

\icmlkeywords{Machine Learning, Reinforcement Learning, Markov Decision Process, Tensor Networks, Matrix Product States, Conditioned Dynamics}

\vskip 0.3in
]

\printAffiliationsAndNotice{}

\begin{abstract}
Tensor network (TN) techniques - often used in the context of quantum many-body physics - have shown promise as a tool for tackling machine learning (ML) problems. The application of TNs to ML, however, has mostly focused on supervised and unsupervised learning. Yet, with their direct connection to hidden Markov chains, TNs are also naturally suited to Markov decision processes (MDPs) which provide the foundation for
reinforcement learning (RL). Here we introduce a general TN formulation of finite, episodic and discrete MDPs. We show how this formulation allows us to exploit algorithms developed for TNs for policy optimisation, the key aim of RL. As an application we consider the issue - formulated as an RL problem - of finding a stochastic evolution that satisfies specific dynamical conditions, using the simple example of random walk excursions as an illustration.
\end{abstract}

\section{Introduction}

In recent years, machine learning methods have found increasing application in various branches of physics, such as the detection of phase transitions \cite{Torlai2016,Rem2019} or variational methods for quantum theory \cite{Nagy2019,Vicentini2019,Hartmann2019,Yoshioka2019}. Similarly, techniques and concepts of physics have found application in machine learning. A fruitful example are tensor networks (TNs) \cite{Montangero2018}. Originally applied in quantum many body physics as variational approaches for ground-state approximation, TNs provide a way to efficiently parametrise important low-dimensional manifolds -- such as those with short-range correlations -- in otherwise unmanageably high-dimension spaces \cite{Eisert2010,Eisert2013,Brandao2015,Huang2019}. Together with a powerful set of optimisation algorithms - known broadly as density-matrix-renormalisation-group (DMRG) \cite{White1992,White1993,Schollwock2011} - TN methods have become state-of-the-art in several areas of physics, and continue to develop rapidly in others \cite{Orus2019}.

Given the nature of the problems tackled by TNs in physics, it is natural to consider them for machine learning problems, where similar issues of finding and optimising efficient parametrisations of relevant manifolds in high-dimensional spaces are key. Indeed, TNs have been applied to image classification \cite{Stoudenmire2016,Sun2019,Efthymiou2019}, unsupervised learning \cite{Han2018,Stoudenmire2018}, deep learning \cite{Levine2019,Gao2019} and probabilistic graphical models \cite{Glasser2018,Glasser2019}.

Despite this rapid progress, little has been done in applying TNs to reinforcement learning (RL) \cite{Sutton2018}, one of the largest fields of machine learning.
Utilised for the solution of games \cite{Mnih2015,Silver2016} or the training of robots \cite{Schulman2015,Haarnoja2018}, RL is often modelled using the formalism of Markov decision processes (MDPs), consisting of repeated updates according to an agent's decision making policy and the dynamics of an environment it inhabits.
Due to the clear product structure of the trajectory probabilities generated by an MDP it is natural to frame such problems as TNs.
Given that TNs are typically applied to problems of numerical optimisation, e.g. via DMRG, this alternative perspective could lead to novel approaches to policy optimization, or suggest useful structures of function approximations. 

Here, we introduce a TN formulation of MDPs - specifically a representation of the expected return in finite MDPs (FMDPs) - and consider how a simple DMRG inspired algorithm can be applied to policy optimisation. 
Since MDPs are closely related to the conditioned stochastic dynamics \cite{Majumdar2015}, already treated with the TN formalism \cite{Garrahan2016}, we use this setting to illustrate the construction and the corresponding optimisation algorithm.
As a specific example we consider an elementary problem of conditioned stochastic dynamics, the generation of stochastic excursions, which can be phrased as an FMDP and solved exactly using the DMRG method introduced.

The layout of the paper is as follows: In Sect. \ref{sect:MDP_Def}, after briefly discussing FMDPs, we outline the TN formalism applied to some relevant examples from stochastic dynamics such as hidden Markov models (HMMs) and the representation of time-integrated observables. The TN formalism for the FMDP is described in Sect. \ref{sect:TNR_FMDP}, while in Sect. \ref{sect:DMRG_opt} we discuss how a DMRG type algorithm can be used to solve policy optimisation numerically. This is illustrated in Sect. \ref{sect:Conditioned_Dyn} with an application to conditioned dynamics. We conclude in Sect. \ref{sect:conclusions} discussing a number of possible extensions where simplifying features of the cases considered are lost.
 
\section{Finite Markov Decision Processes and Tensor Networks for Dynamics}
\label{sect:MDP_Def}

\subsection{Finite Markov Decision Processes}
In a discrete-time, episodic, MDP, individual trajectories take the form,
\begin{align}
\text{Traj} = \left( S_{0}, A_{0}, R_{1}, S_{1}, A_{1}, R_{2}, S_{2}, A_{2}, ..., R_{T}, S_{T} \right) ~ ,
\end{align}
where $S_{t}, A_{t}$ and $R_{t}$ are random variables for the state, action and reward, taking values $s_{t}, a_{t}$ and $r_{t}$ respectively. The termination time, $T$, can also be a random variable, though we consider it fixed for simplicity. At $T$, when the episode ends, the state of the system is a terminal state. We will assume this to be unique and denote it $s^{+}$, so that $S_{T} = s^{+}$ for all trajectories. We will further assume that all random variables take on values from ($t$-independent) finite sets, $\mathcal{S}, \mathcal{A}$ and $\mathcal{R}$. We will call this scenario an FMDP.

In an FMDP, the probabilities dictating the dynamics are:
\begin{align}
\text{Pr}\lbrace S_{t} = s' , R_{t} = r | S_{t-1} = s, A_{t-1} = a\rbrace = p_{t}(s', r | s, a) ~,
\nonumber
\end{align}
where $t = 1, ..., T$ and $p_{t} : \Ss \times \Rs \times \Ss \times \As \to \left[0,1\right]$ is the function defining the dynamics from $t-1 \to t$. This obeys the normalisation condition,
\begin{align}
\sum_{s',r} p_{t}(s',r | s, a) = 1 ~ \forall ~ s, a, t ~ .
\label{eqn:norm_p}
\end{align}

At the beginning of the episode, the initial state distribution is given by,
\begin{align}
\text{Pr}\lbrace S_{0} = s \rbrace = p_{0}(s) ~.
\end{align}

At termination time, when the episode ends:
\begin{align}
& \text{Pr}\lbrace S_{T} = s' , R_{T} = r | S_{T-1} = s, A_{T-1}  = a\rbrace  ~  = \nonumber \\
& = p_{T}\left(s',r | s,a \right) ~ , \\
& = p_{T}(s^{+}, r | s, a) \delta_{s',s^{+}} ~.
\label{eqn:terminal_dynamics}
\end{align}

The decision component of the FMDP is contained in the probability of selecting a particular action conditioned on a given state,
\begin{align}
\text{Pr}\lbrace A_{t-1} = a | S_{t-1} = s\rbrace = \pi_{t}(a|s) ~,
\end{align}
where the function $\pi_{t} : \As \times \Ss \to [0,1]$ is known as the policy (labelled by $t = 1, 2, ..., T$), and is normalised,
\begin{align}
\sum_{a} \pi_{t}(a | s) = 1 ~ \forall ~ s, ~ t ~.
\label{eqn:norm_pi}
\end{align}

The return of an episode, $G_{1:T}$, is defined as,
\begin{align}
G_{t_{I}:t_{F}} = \sum_{t=t_{I}}^{t_{F}} R_{t} ~ .
\end{align}
An optimal policy, $\bm{\pi}^{*} = (\pi^{*}_{1}, ..., \pi^{*}_{T})$, is then a policy that maximises the expected return,
\begin{align}
\bm{\pi}^{*}  = \text{argmax}_{\pi_{1}, \pi_{2}, ..., \pi_{T}} \mathds{E}\left[ G_{1:T} \right] ~ .
\end{align}

Policy optimisation for a given FMDP is thus a high-dimensional constrained optimisation problem: The total number of parameters is $T \times \dSs \times \dAs$, and each $\pi_{t}$ is constrained to lie on the manifold of stochastic matrices, i.e. they obey Eq. \eqref{eqn:norm_pi} and have elements in $[0,1]$. 

\subsection{Tensor Networks for Hidden Markov Models}

A TN is a collection of tensors contracted together in a given pattern, typically specified by a graph. An elementary example of this is a chain of matrices $\mathbf{M}_i$ applied to a vector $\ket{p_{0}}$ written in the braket notation common to physics,
\begin{align}
\mathbf{M}_{T} ... \mathbf{M}_{2}\mathbf{M}_{1} \ket{p_{0}} ~,
\label{eqn:Markov_Chain}
\end{align}
where $\ket{p_{0}}=\sum_s c_s\ket{s}$ for some coefficients $c_s$ and the vectors $\ket{s}$ associated to each state $s$ form a basis of a vector space $\vS$.
Since the matrices are rank-$2$ tensors and the vector a rank-$1$ tensor, this can be considered a TN consisting of $T+1$ tensors, where the contraction pattern is given by the usual matrix products. Performing such a contraction produces a new vector, $\ket{p_{T}}$, with components,
\begin{align}
\braket{s_{T}|p_{T}} = \braket{s_{T}|\mathbf{M}_{T} ... \mathbf{M}_{2}\mathbf{M}_{1} | p_{0}} ~,
\label{eqn:Markov_chain_comp}
\end{align}
where $\bra{s_T}$ are dual basis vectors such that the dot product $\left\langle s | s' \right\rangle=\delta_{ss'}$.
In this sense, we can say that this product is a TN representation (TNR) of the vector $\ket{p_{T}}$.

While the TN structure of Eq. \eqref{eqn:Markov_chain_comp} is simple and requires no clarification, more generally it is convenient to specify the contraction pattern of a TN via a graph, using a standard diagrammatic notation. In such a notation, rank-$K$ tensors are represented as shapes with $K$ legs, and contractions are indicated by joining the appropriate legs together. In this notation, Eq. \eqref{eqn:Markov_chain_comp} for $T=3$ reads,
\begin{align}
\includegraphics[width=0.8\linewidth]{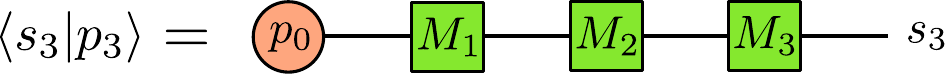}
\end{align}

As suggested by the chosen notation, Eq. \eqref{eqn:Markov_chain_comp} is exactly the TNR for a probability distribution over states produced by a Markovian dynamics. In that case, the components of $\mathbf{M}_{t}$ are equal to the probabilities of state transitions,
\begin{align}
\braket{s_{t+1}|\mathbf{M}_{t}|s_{t}} = p_{t}\left(s_{t+1} | s_{t}\right) ~ ,
\end{align} 
while the components of $\ket{p_{0}}$ give the initial probability distribution over states. Any vector $\ket{p_{t}} \in \vS$, that is a convex combination of the particular basis $\ket{s}$, can be interpreted as a probability distribution over states via their components $\braket{s|p_{t}} = p_{t}(s)$. This implies their normalisation as $\braket{-_{s} | p_{t}} = 1$, where $\bra{-_{s}}$ is the flat-vector for the basis, $\bra{-_{s}} = \sum_{s} \bra{s}$. The matrices, $\mathbf{M}_{t}$, which can be considered as elements of $\vS \otimes \vS^{*}$, obey a related condition,
\begin{align}
\bra{-_{s}}\mathbf{M}_{t} = \bra{-_{s}} ~~ \forall ~~ t ,
\end{align}
which allows for the interpretation of their components as conditional probabilities.

A more complicated TNR relevant for dynamics is offered by the matrix product state (MPS) representation - also know as the tensor train decomposition \cite{Oseledets2011} -  of HMMs. In a system described by an HMM, the dynamics is Markovian and thus governed by $p_{t}(s'|s)$ or $\mathbf{M}_{t}$. However, information about the state, $\mathbf{s} = \left(s_{1}, s_{2}, ..., s_{T} \right)$, cannot be accessed directly, and only partial information is revealed through the observables at each time step. We will denote these observables as $\mathbf{r} = \left(r_{1}, r_{2}, ..., r_{T}\right)$, since they will correspond to the rewards in the FMDPs considered later. 

In HMMs, the relevant probabilities are then $p_{t}\left(r,s'|s\right)$, which are related to $p_{t}\left(s' | s\right)$ through marginalisation;
\begin{align}
p_{t}\left(s' | s\right) = \sum_{r} p_{t}\left(s',r|s\right) ~ .
\label{eqn:reward_marginalisation}
\end{align}
The marginalisation Eq. \eqref{eqn:reward_marginalisation}, can also be viewed as a decomposition of the matrices $\mathbf{M}_{t}$, corresponding to $p_{t}(s'|s)$, into a sum of matrices $\mathbf{M}^{r}_{t}$, with components $p_{t}\left(s',r|s\right)$,
\begin{align}
\mathbf{M}_{t} = \sum_{r} \mathbf{M}_{t}^{r} ~.
\label{eqn:MD_HMM_Decomp}
\end{align}
This decomposition implies the introduction of an encoding for the possible observations, $r \to \ket{r}$, where $r \in \mathcal{R}$ and $\ket{r} \in \vR$, which can be achieved in the same way as for states. Taking tensor products of these vectors produces an encoding for the possible observations over time, $\ket{\mathbf{r}}$, which form a basis of the vector space $\vR^{T} = \bigotimes_{t=1}^{T} \vR $. The vectors representing the probability distributions over observations are elements of this space, $\ket{p_{\mathbf{r}}} \in \vR^{T}$, and obey the normalisation
$\braket{-_{\mathbf{r}} | p_{\mathbf{r}}} = 1$.

To build a TNR for the HMM, one can consider the set of matrices, $\mathbf{M}^{r}_{t}$, as a rank-$3$ tensors, which are elements of $\vS \otimes \vR \otimes \vS^{*}$. The decomposition, Eq. \eqref{eqn:MD_HMM_Decomp} can then be expressed as an equation relating tensors. Graphically, representing the flat-vector as a vertical line, Eq. \eqref{eqn:MD_HMM_Decomp} reads,
\begin{align}
\includegraphics[width=0.8\linewidth]{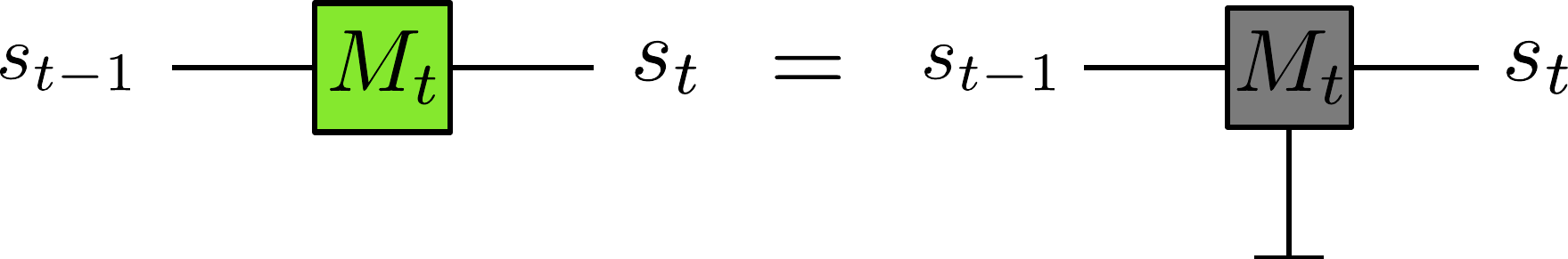}
\label{eqn:HMM_MPS_Decomp}
\end{align}
Since the flat-vector, $\ket{-_{r}}$, causes the marginalisation of $r$, one can remove the marginalisation by removing this vector. Thus, one finds that the probability of a making a particular set of observations in an HMM can then be expressed as,
\begin{align}
\braket{\mathbf{r}|p_{\mathbf{r}}} = \braket{-_{s_{T}}|\mathbf{M}_{T}^{r_{T}}...\mathbf{M}_{2}^{r_{2}}\mathbf{M}_{1}^{r_{1}}|p_{0}} ~ .
\end{align}
This is exactly the structure of an MPS \cite{Schollwock2011}. For $T=3$, the TNR of $\ket{p_\mathbf{r}}$ in a HMM is thus,
\begin{align}
\includegraphics[width=0.8\linewidth]{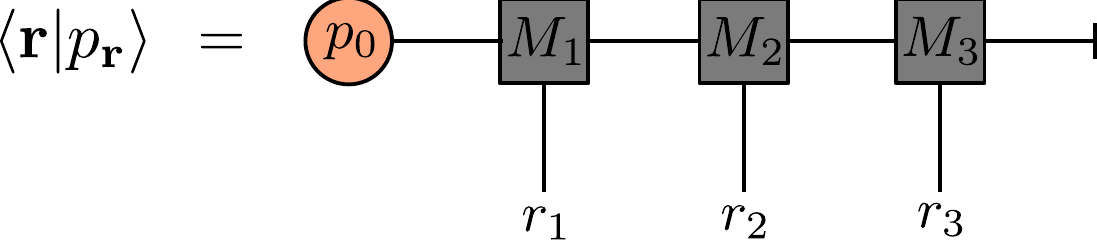}
\label{eqn:HMM_MPS}
\end{align}

\subsection{Tensor Networks for Time Integrated Observables}

When considering dynamics described by an HMM, one is often interested in time integrated observables. Such objects can be represented easily in terms of TNs, which in turn allows for the TNR of averages or higher-order moments. 

To specialise to the case of MDPs, we will consider a HMM where we observe a reward at each discrete time-step, $t = 1,\ldots,T$, and wish to represent the expected return, $\mathds{E}\left[G_{1:T}\right]$ as a TN. To achieve this, one begins by introducing the operator, $\hat{R}$, which is diagonal in the encoding basis,
\begin{align}
\hat{R}\ket{r} &= r\ket{r} ~ ,
\end{align}
such that it can be used to produce the explicit values observed at a given time.

To encode the return of an episode, $G_{1:T}$, one defines the operator, $\hat{G}_{1:T}$, acting on the vector space spanned by $\ket{\mathbf{r}}$,
\begin{align}
\hat{G}_{1:T} = \sum_{t=1}^{T} \hat{R}_{t} ~ ,
\end{align}
where  $\hat{R}_{t}$ is the operator acting as $\hat{R}$ on the $t^{\text{th}}$ vector space in the tensor product or as identity otherwise. 

An appropriate TNR for $\hat{G}_{1:T}$ is offered by the matrix product operator (MPO) \cite{PerezGarcia2007,Crosswhite2008,Pirvu2010}. The MPO is defined analogously to the MPS,
\begin{align}
\braket{\mathbf{r'}|\hat{G}_{1:T}|\mathbf{r}} &= \mathbf{w}_{1}^{r_{1},r_{1}'}\mathbf{W}_{2}^{r_{2},r_{2}'}...\mathbf{w}_{T}^{r_{T},r_{T}'} ~.
\label{eqn:episode_return_mpo}
\end{align}
This has the same contraction pattern as the MPS, as is clear from the corresponding graphical equation, e.g.,
\begin{align}
\includegraphics[width=0.7\linewidth]{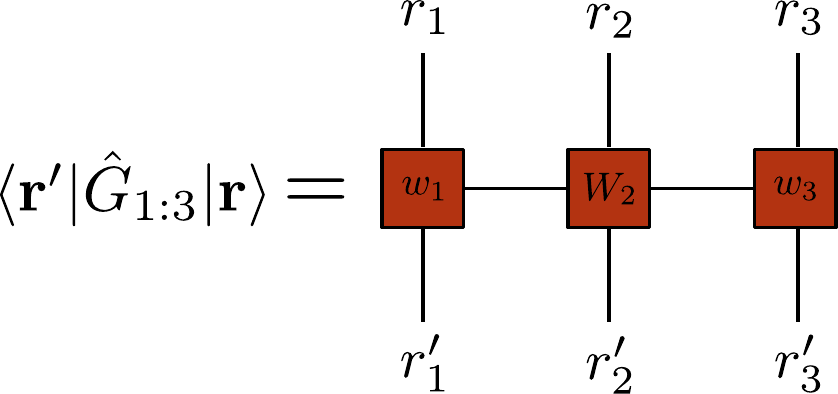}
\end{align}

The sets of matrices, $\mathbf{W}_{t}^{r_{t},r_{t}'}$, can be chosen using standard construction methods \cite{Schollwock2011}: Consider the operator-valued matrix,
\begin{align}
\mathbf{W}_{t} = \begin{pmatrix}
\mathds{1} & 0 \\ 
\hat{R} & \mathds{1}
\end{pmatrix} ~ .
\end{align}
When multiplied, components of such matrices are to be combined via tensor products:
\begin{align}
\mathbf{W}_{t}\mathbf{W}_{t+1} &= \begin{pmatrix}
\mathds{1} & 0 \\ 
\hat{R} & \mathds{1}
\end{pmatrix} \begin{pmatrix}
\mathds{1} & 0 \\ 
\hat{R} & \mathds{1}
\end{pmatrix} ~ , \\
&= \begin{pmatrix}
\mathds{1}\mathds{1} & 0 \\ 
\hat{R}\mathds{1} + \mathds{1}\hat{R} & \mathds{1}\mathds{1}
\end{pmatrix} ~ .
\end{align}
Further defining the operator-valued boundary-vectors:
\begin{align}
\mathbf{w}_{1} = \begin{pmatrix}
\mathds{1} \\ 
\hat{R}
\end{pmatrix} ~ ,
\end{align}
\begin{align}
\mathbf{w}_{T} = \begin{pmatrix}
\hat{R} & 
\mathds{1}
\end{pmatrix} ~ .
\end{align}
The return operator for the whole episode, $\hat{G}_{1:T}$, can be represented as product of such matrices,
\begin{align}
\hat{G}_{1:T} &= \mathbf{w}_{1}\mathbf{W}_{2}...\mathbf{w}_{T} ~ .
\end{align}

In the basis $\ket{\mathbf{r}}$, this corresponds to the previously defined MPO form \eqref{eqn:episode_return_mpo} with matrices:
\begin{align}
\mathbf{w}_{1}^{r_{1},r_{1}'} = \begin{pmatrix}
\delta_{r_{1},r_{1}'} \\ 
r_{1} \delta_{r_{1},r_{1}'} 
\end{pmatrix} ~ ,
\end{align}
\begin{align}
\mathbf{w}_{T}^{r_{T},r_{T}'} = \begin{pmatrix}
r_{T} \delta_{r_{T},r_{T}'} & 
\delta_{r_{T},r_{T}'}
\end{pmatrix} ~ ,
\end{align}
\begin{align}
\mathbf{W}_{t}^{r_{t}} = \begin{pmatrix}
\delta_{r_{t},r_{t}'} & 0 \\ 
r_{t} \delta_{r_{t},r_{t}'}  & \delta_{r_{t},r_{t}'}
\end{pmatrix} ~ .
\end{align}

With this definition, the TNR of the expected return can be obtained directly from its expression in terms of operators and vectors,
\begin{align}
\mathds{E}[G_{1:T}] = \braket{-_{\mathbf{r}}|\hat{G}_{1:T}|p_{\mathbf{r}}} ~ .
\label{eqn:encoded_expected_return}
\end{align}
Contracting together the constituent TNRs of $\ket{p_{\mathbf{r}}}$, $\bra{-_{\mathbf{r}}}$ and $\hat{G}_{1:T}$ gives the overall TN expression. For example, if $T=3$ this is given by,
\begin{align}
\includegraphics[width=0.8\linewidth]{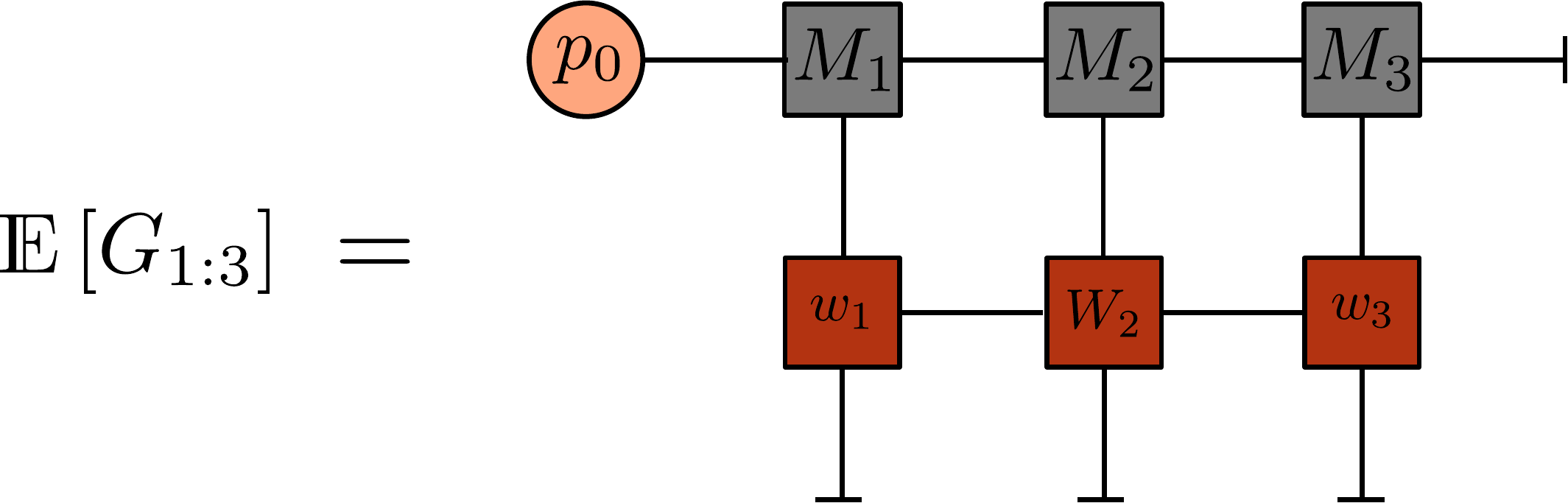}
\label{eqn:TNR_expected_return}
\end{align}

Similarly, representations of higher-order observables such at $\mathds{E}[G_{1:T}^{2}]$ can be constructed directly from,
\begin{align}
\mathds{E}[G_{1:T}^{2}] = \braket{-_{\mathbf{r}}|\hat{G}_{1:T}\hat{G}_{1:T}|p_{\mathbf{r}}} ~ ,
\label{eqn:encoded_expected_square_return}
\end{align}
as can TNRs of any other observables for which there are MPO representations of the relevant operators.

\section{Tensor Network Representations for FMDPs}
\label{sect:TNR_FMDP}

As with HMMs, the expected return of an FMDP can be expressed as a TN by using an MPS representation of $\ket{p_{\mathbf{r}}}$ and an MPO representation of $\hat{G}_{1:T}$. However, in order to perform policy optimisation using the tools developed for TNs, the dependence of $\mathds{E}\left[G_{1:T}\right]$ on the policy, via that of $\ket{p_{\mathbf{r}}}$, must be extracted explicitly. 

Similar to when moving from Markovian dynamics to HMMs, this can be achieved by relating the relevant probabilities in the FMDP to those of the HMM, $p_{t}\left(s_{t},r_{t}|s_{t-1}\right)$, via marginalisation;
\begin{align}
p_{t}(s_{t},r_{t} | s_{t-1}) = \sum_{a_{t-1}} p_{t}(s_{t},r_{t} | s_{t-1}, a_{t-1}) \pi_{t}(a_{t-1} | s_{t-1}) ~ .
\label{eqn:action_marginalisation}
\end{align}
By expressing this as a relationship between tensors, one can extend the TNR of $\mathds{E}\left[G_{1:T}\right]$ used for HMMs, Eq. \eqref{eqn:TNR_expected_return}, to FMDPs. 

To begin, we rewrite Eq. \eqref{eqn:action_marginalisation} as,
\begin{align}
\braket{s_{t}|\mathbf{M}_{t}^{r_{t}}|s_{t-1}} = \sum_{a_{t-1}} \braket{s_{t}|\mathbf{M}_{t}^{r_{t},a_{t-1}}|s_{t-1}} \pi^{a_{t-1},s_{t-1}}_{t} ~,
\label{eqn:HMM_Decomp}
\end{align}
where $\mathbf{M}^{r,a}$ are sets of matrices labelled by both a reward $r$ and an action $a$. As for $s$ and $r$, an encoding of $a \in \mathcal{A}$ into vectors is implied. Note that, because the index with value by $s_{t-1}$ appears twice but is not summed over (i.e. it is not part of a contraction), Eq. \eqref{eqn:HMM_Decomp} does not yet provide a TN decomposition of Eq. \eqref{eqn:action_marginalisation} as desired, as can be seen clearly from the graphical notation;
\begin{align}
\includegraphics[width=0.8\linewidth]{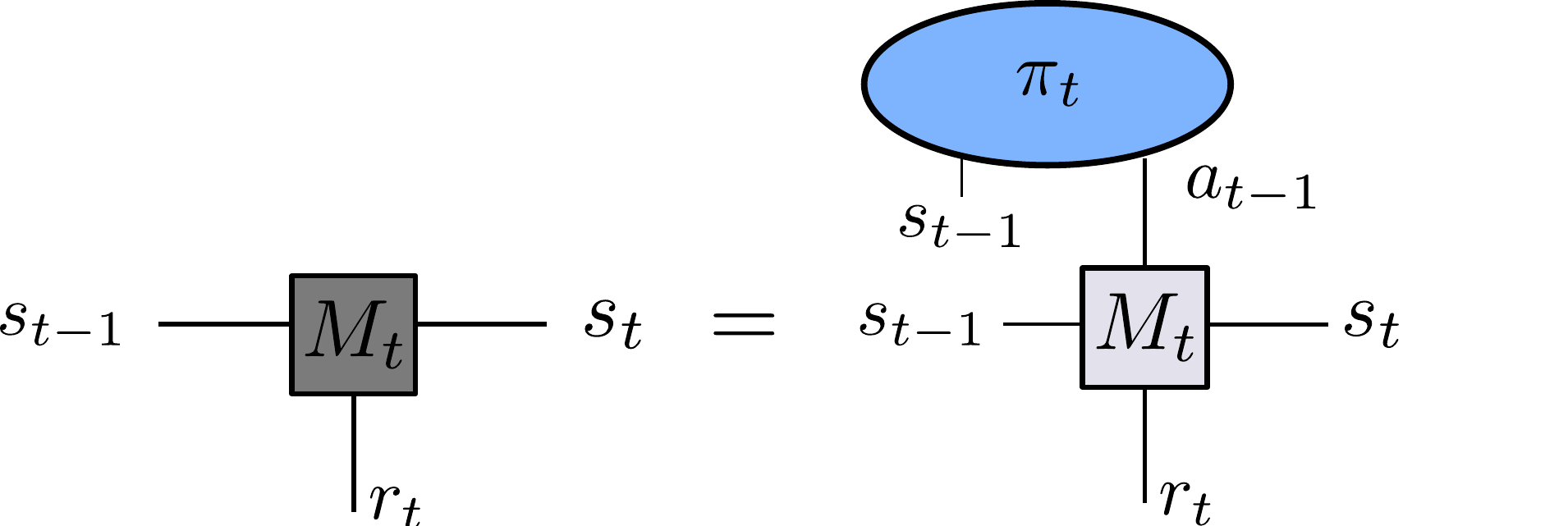}
\end{align}
Note that while for convenience we have use the same tensor label, $M_{t}$ for both the rank-$3$ and rank-$4$ tensors, these are distinct objects as indicated by the different colours. 

To express the decomposition Eq. \eqref{eqn:HMM_Decomp} in terms of tensors and their contractions alone, one must account for the fact that the information about the state at the previous time-step, $s_{t-1}$, is used for conditioning twice; once in the policy and once in the dynamics. In TNs, such additional conditioning requires the inclusion of copy tensors \cite{Biamonte2011,Glasser2018}. Defined with respect to a chosen basis, $\ket{s}$, the components of a copy tensor have value one if all indices are equal, and zero otherwise. The copy tensor we will consider is the rank-$3$ copy tensor, whose components can be defined as a set of matrices,
\begin{align}
\mathbf{\Delta}^{s} = \ket{s}\hspace{-2.1pt}\bra{s}~ .
\end{align}
For example,
\begin{align}
\braket{s'|\mathbf{\Delta}^{0}|s''} = \braket{s'|0}\hspace{-2.1pt}\braket{0|s''} = \delta_{0,s'}\delta_{0,s''} ~.
\end{align}
Graphically, we denote the copy tensor as a black circle,
\begin{align}
\includegraphics[width=0.45\linewidth]{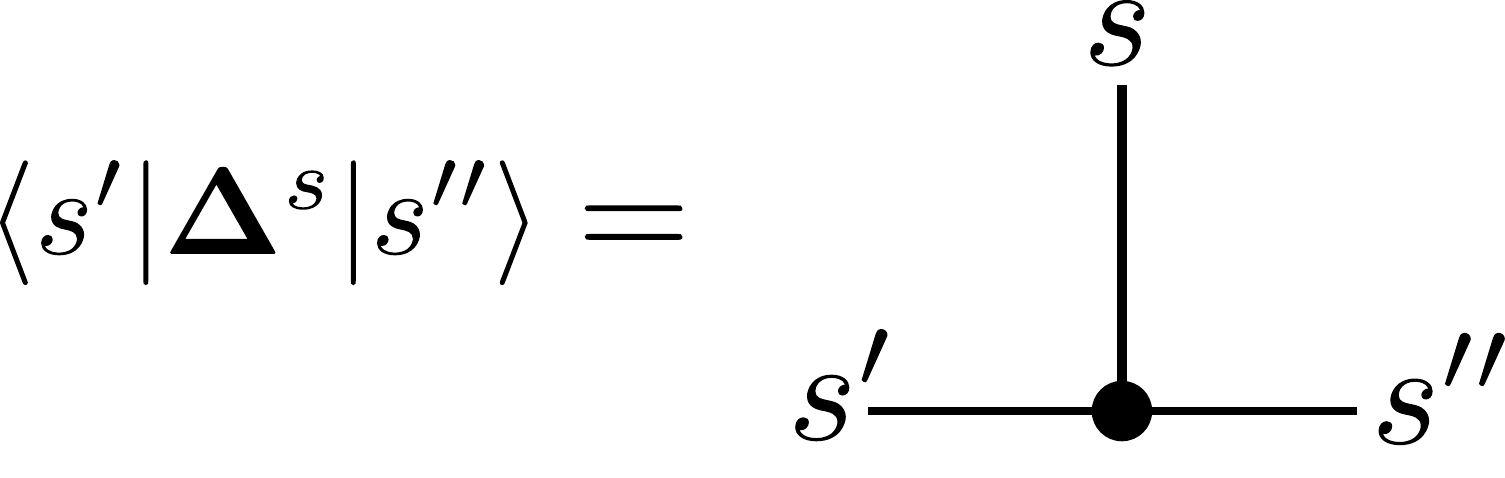}
\end{align}
In general, including multiple copy tensors will allow the construction of a TNR for any joint probability distribution via decomposition using the chain rule. In such a TNR, each variable will be associated to a number of copy tensors equal to the number of times it is reused for conditioning. Thus, the TNR for a joint probability distribution decomposed via the chain-rule is in general a two-dimensional, hierarchical TN. In an FMDP, this structure is simplified considerably by the Markovian assumption, so that only a single state-copy tensor is required per time-step, and a one-dimensional TNR (an MPS) results.

With the copy tensor defined, the decomposition Eq. \eqref{eqn:HMM_Decomp} can be expressed in terms of tensors as, 
\begin{align}
\braket{s_{t}|\mathbf{M}_{t}^{r_{t}}|s_{t-1}} = \sum_{\underset{a_{t-1}}{\tilde{s}_{t-1}}} \bra{s_{t}}\mathbf{M}_{t}^{r_{t},a_{t-1}} \mathbf{\Delta}^{s_{t-1}}\ket{\tilde{s}_{t-1}} \pi^{a_{t-1},\tilde{s}_{t-1}}_{t} ~ .
\nonumber
\end{align}
In the graphical notation this is,
\begin{align}
\includegraphics[width=0.8\linewidth]{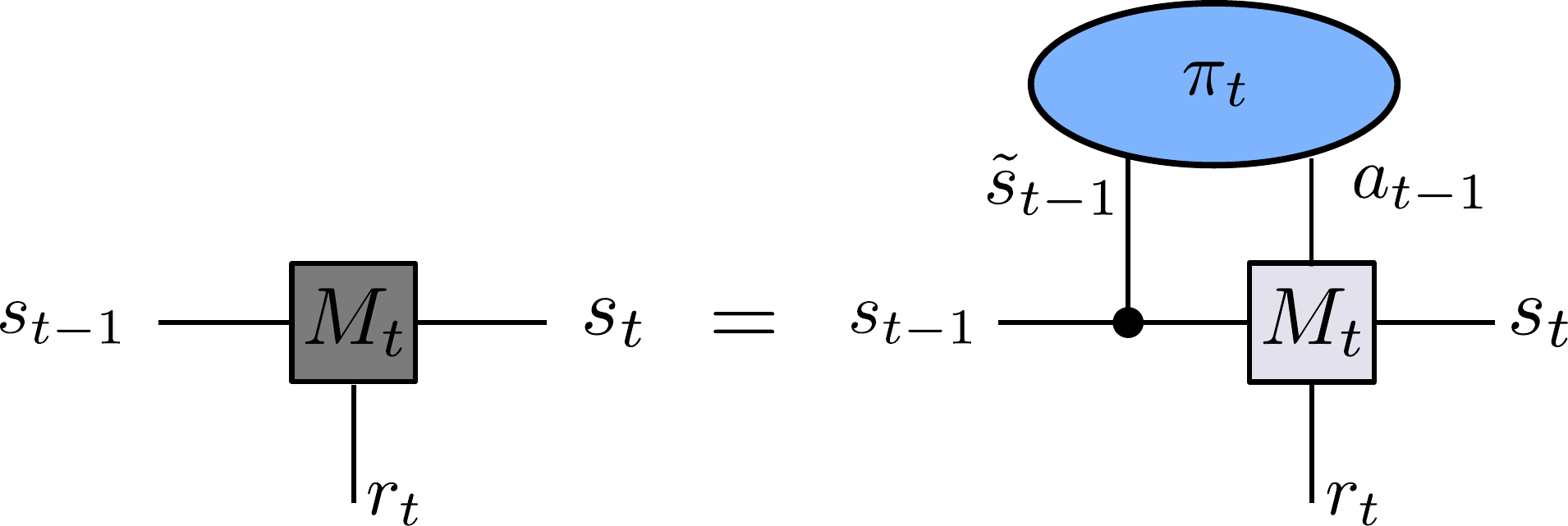}
\end{align}
The notation for this decomposition can be further simplified by grouping the copy-tensor and rank-$4$ matrix together, while also considering the indices corresponding to the state action variables, $\tilde{s}_{t-1}$ and $a_{t-1}$, as a single compound index;
\begin{align}
\braket{s_{t}|\mathbf{M}^{r_{t}}_{t}|s_{t-1}} =& \nonumber  \\
~~ =  \sum_{(\tilde{s}_{t-1},a_{t-1})}  &\braket{s_{t}|\tilde{\mathbf{M}}_{t}^{r_{t} , (\tilde{s}_{t-1},a_{t-1})}|s_{t}} \pi_{t}^{(\tilde{s}_{t-1},a_{t-1})} ~ ,
\nonumber
\end{align}
or,
\begin{align}
\includegraphics[width=0.8\linewidth]{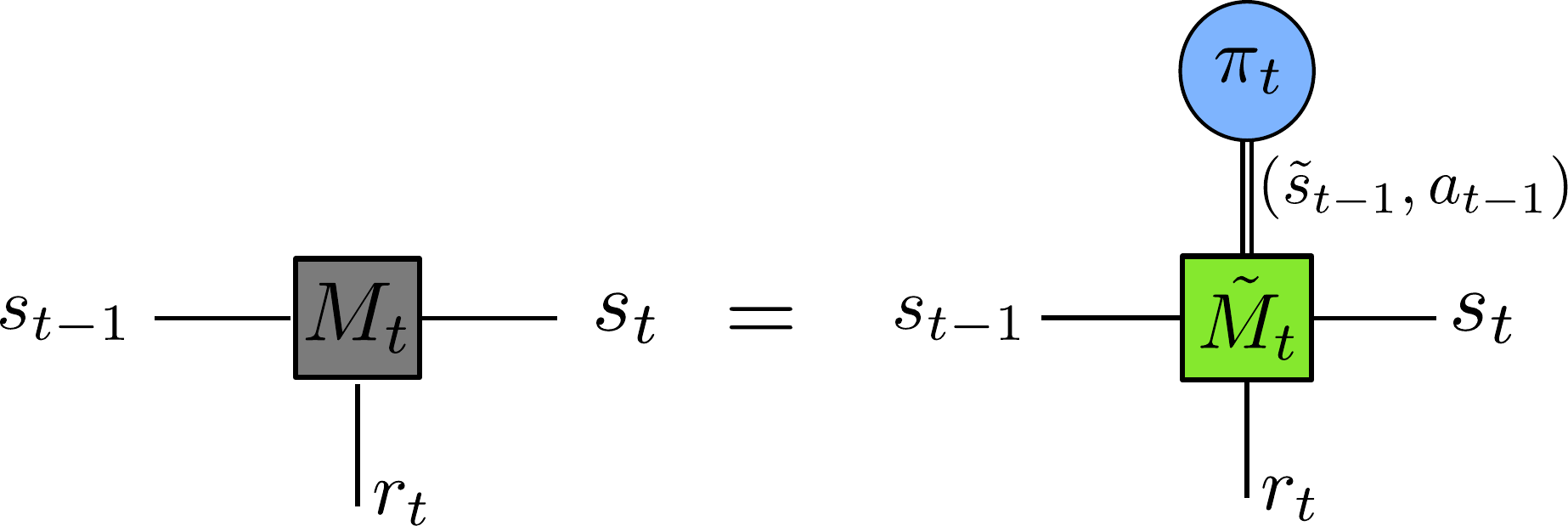}
\end{align}
Written in this way, one can see that the rank-$3$ tensors appearing in the MPS representation of $\ket{p_{\mathbf{r}}}$ can be considered as the contraction of a vector, $\ket{\pi_{t}}$, with components $\pi^{(s,a)}_{t}$, and a rank-$4$ tensor, $\tilde{M}_{t}$. This is exactly the MPS representation that results from the application of an operator, $\hat{M}$, expressed as an MPO, to a vector $\ket{\bm{\pi}}$, expressed as an MPS, i.e. $\ket{p_{\mathbf{r}}} = \hat{M}\ket{\bm{\pi}}$. Graphically, this can be seen by decomposing every tensor in the MPS expression of $\ket{p_{\mathbf{r}}}$, Eq. \eqref{eqn:HMM_MPS}. For example, when $T=3$, this gives,
\begin{align}
\includegraphics[width=0.8\linewidth]{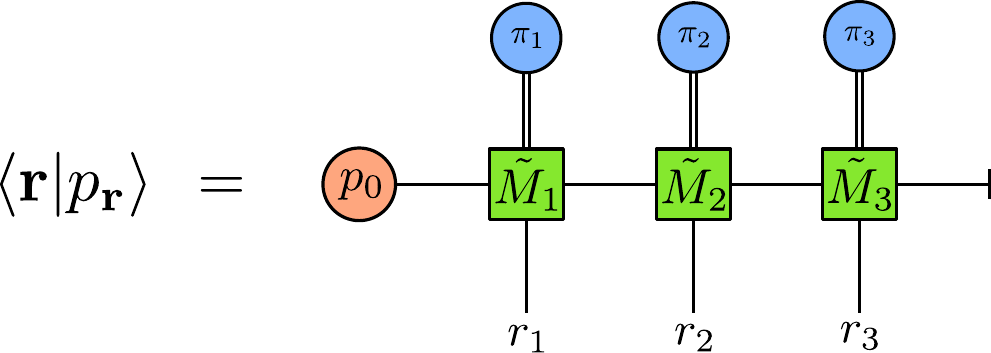}
\end{align}
In this expression, the vector, $\ket{\bm{\pi}} = \ket{\pi_{1}, \pi_{2}, ...,\pi_{T}}$, is represented as a product-state MPS, i.e. an MPS with $\chi = 1$:
\begin{align}
\braket{\mathbf{s},\mathbf{a}|\bm{\pi}} = \pi_{1}^{(s_{0},a_{0})} \pi_{2}^{(s_{1},a_{1})} \times ... \times \pi_{T}^{(s_{T-1},a_{T-1})} ~ ,
\end{align}
or, for $T=3$,
\begin{align}
\includegraphics[width=0.7\linewidth]{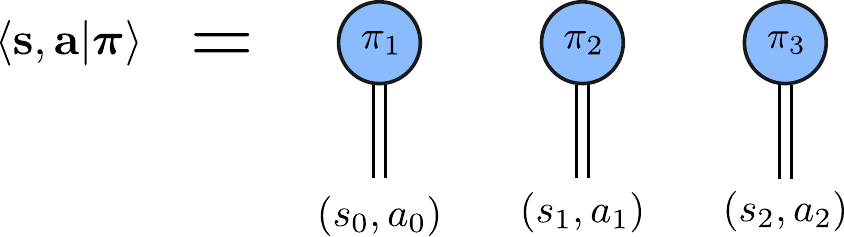}
\end{align}

The vector $\ket{\bm{\pi}}$ contains all information about the policy during the FMDP. Information about the dynamics is instead contained in the operator $\hat{M}$, which has the MPO representation,
\begin{align}
\braket{\mathbf{r}|\hat{M}|\mathbf{s,a}} &=  \tilde{\mathbf{m}}_{1}^{r_{1},(s,a)_{0}}\tilde{\mathbf{M}}_{2}^{r_{2},(s,a)_{1}}...\tilde{\mathbf{m}}_{T}^{r_{T},(s,a)_{T-1}} ~,
\end{align}
where for convenience we have defined $\tilde{\mathbf{m}}_{1}^{r_{1},(s,a)_{0}} = \tilde{\mathbf{m}}_{0}\tilde{\mathbf{M}}_{1}^{r_{1},(s,a)_{0}}$, with $\tilde{\mathbf{m}}_{0}$ encoding the initial probability distribution as, $\left[\tilde{\mathbf{m}}_{0}\right]_{\beta} = \text{Pr}\left[S_{0} = \beta\right] = p_{0}(\beta) ~$, and $\tilde{\mathbf{m}}_{T}^{r_{T},(s,a)_{T-1}} = \tilde{\mathbf{M}}_{T}^{r_{T},(s,a)_{T-1}}\ket{-_{s}}$. 

With the decomposition $\ket{p_{\mathbf{r}}} = \hat{M}\ket{\bm{\pi}}$, the desired TNR of the expected return is complete. For example, when $T=3$, the expected return can be written graphically as,
\begin{align}
\includegraphics[width=0.7\linewidth]{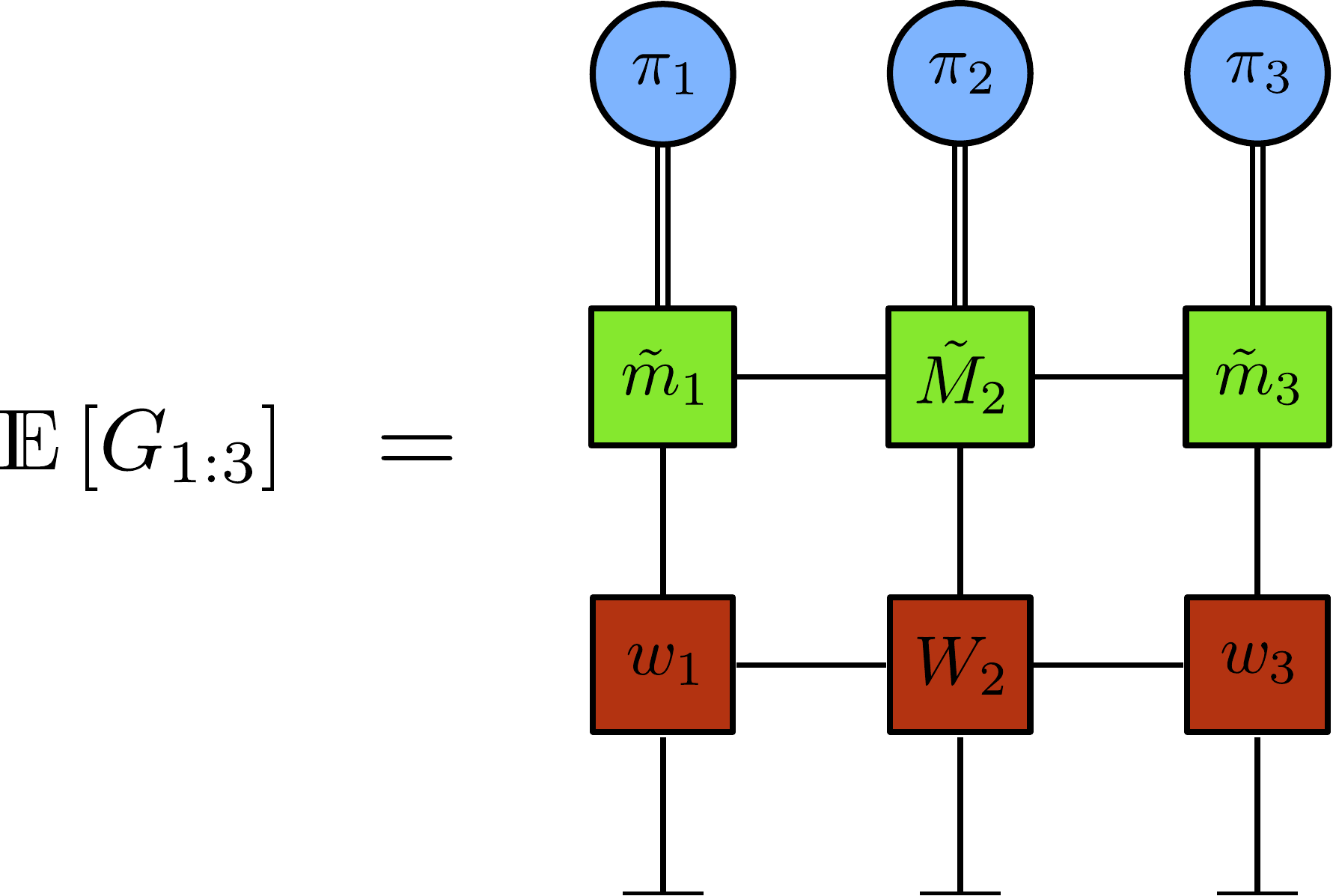}
\end{align}

\begin{figure*}[t]
\centering
\includegraphics[width=0.8\linewidth]{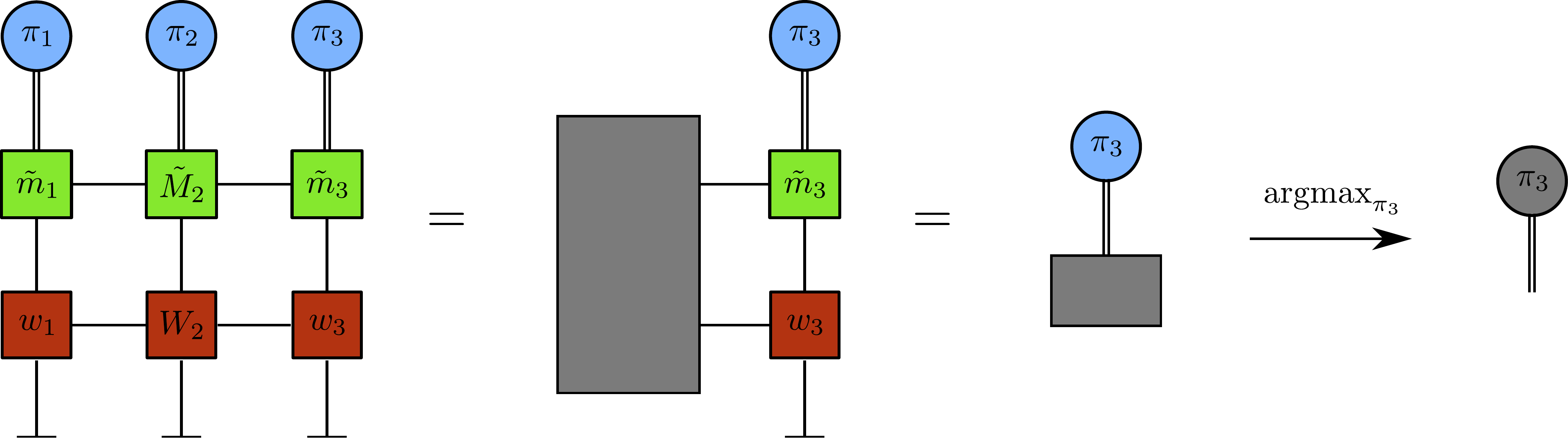}
\caption{\textbf{DMRG style policy optimisation:} With the expected return expressed as an TN, one can perform policy optimisation by considering each tensor of $\ket{\bm{\pi}}$ in turn. At each tensor -- taking here $\pi_{3}$ as an example -- one contracts the rest of the network (the environment) which is considered fixed at this iteration. The optimal tensor is then found with respect to this environment. This is used to update the policy and thus is subsequently used to calculate a new environment for the next iteration of the algorithm. The cost of the contraction is dominated by the size of the set of states, and scales as $\mathcal{O}\left(|\mathcal{S}|^{3}\right)$.}
\label{fig:DMRG_opt}
\end{figure*}

\section{DMRG Approach to Policy Optimisation}
\label{sect:DMRG_opt}

With an objective function expressed as a TN, an approach to optimisation that has proved effective is to optimise just one or two tensors at a time, while keeping the other tensors - the ``environment'' - fixed. (Note that this is distinct from what is commonly called the environment in RL, which we refer to as the system dynamics.) By passing (sweeping) through the tensors that contain the variational parameters, one can perform the optimisation iteratively. This allows for the efficient use of computational resources and has been very successful in the context of one dimension quantum many body systems, where is it known as DMRG. Building on this basic idea, a large variety of techniques have been developed to perform sophisticated, state-of-the-art optimisations.   

In the case of policy optimisation using a TNR of the expected return, 
an approach inspired by DMRG \cite{Schollwock2011} can be used, see Fig. \ref{fig:DMRG_opt}. In such an approach, each tensor of $\ket{\bm{\pi}}$ is visited in turn. At a given tensor, labelled by $t$, the expected reward is calculated as a linear map onto this tensor by evaluating the environment - i.e. contracting together all other tensors in the network that are considered fixed at this iteration of the optimisation. In the case we are considering, the environment can be evaluated exactly, though in general approximations are required. The policy is then updated by finding the tensor that maximises the return for this fixed environment, subject to the desired constraints that the tensor $\pi_{t}$ be normalised appropriately, $\sum_{a} \pi_{t}^{s,a} = 1 ~ \forall ~ s$ and have components $\pi_{t}^{s,a} \in \left[0,1\right]$. 

In typical DMRG applications, a back-and-forth sweeping pattern is used to optimise the tensors. While in general many sweeps might be required to reach convergence, in the simple set up we consider a single DMRG sweep backwards in time is sufficient to find the optimal policy exactly. 

To see this, consider splitting the expected return at time $t$: $\mathds{E}\left[G_{1:T}\right] = \mathds{E}\left[G_{1:t}\right] + \mathds{E}\left[G_{t+1:T}\right] ~.$ The first of these terms, $\mathds{E}\left[G_{1:t}\right]$, can be calculated using only the probability distribution over the first $t$ rewards, $\text{Pr}\lbrace R_{1} = r_{1}, R_{2} = r_{2}, ..., R_{t} = r_{t} \rbrace$. Decomposing this probability with the chain-rule forwards in time - i.e. conditioning occurs only on past states and actions - allows for the corresponding expectation value to be computed starting from the initial state distribution, $p_{0}\left(s_{0}\right)$, which is assumed known and policy-independent, using only the policy up to time $t$. The second of these terms, $\mathds{E}\left[G_{t+1:T}\right]$, instead requires $\text{Pr}\lbrace R_{t+1} = r_{t+1}, R_{t+2} = r_{t+2}, ..., R_{T} = r_{T} \rbrace$. Again decomposing this forwards in time the expected value of the return can be calculated. However, in this case the initial (marginal) distribution over states is $p_{t}\left(s_{t}\right)$. Calculating this initial distribution requires knowledge of the policy until time $t$, and thus the policy of the whole episode is required.

Defining $\pi_{t_{I}:t_{F}}$ as a vector containing all the parameters of the policy for $t \in \left[t_{I}, t_{F}\right]$, the dependence on the policy of the two terms in the above decomposition implies that the optimal policy satisfies the simultaneous equations:
\begin{align}
\frac{\partial}{\partial\pi_{1:t}} \mathds{E}\left[G_{1:t}\right] + \frac{\partial}{\partial\pi_{1:t}} \mathds{E}\left[G_{t+1:T}\right] &= 0 \\
\frac{\partial}{\partial\pi_{t+1:T}} \mathds{E}\left[G_{t+1:T}\right] &= 0  ~.
\end{align}
The optimal policy can therefore be found by first solving the second equation, thus finding $\pi_{t+1:T}^{*}$, and then substituting into the first one and solving for $\pi_{1:t}^{*}$. Since the choice of $t$ was arbitrary in the above argument, one can proceed recursively starting from $t=T-1$. This is the usual one-site DMRG type algorithm, starting from the right-most site.

\begin{figure*}[t]
  \centering
    \includegraphics[width=0.8\linewidth]{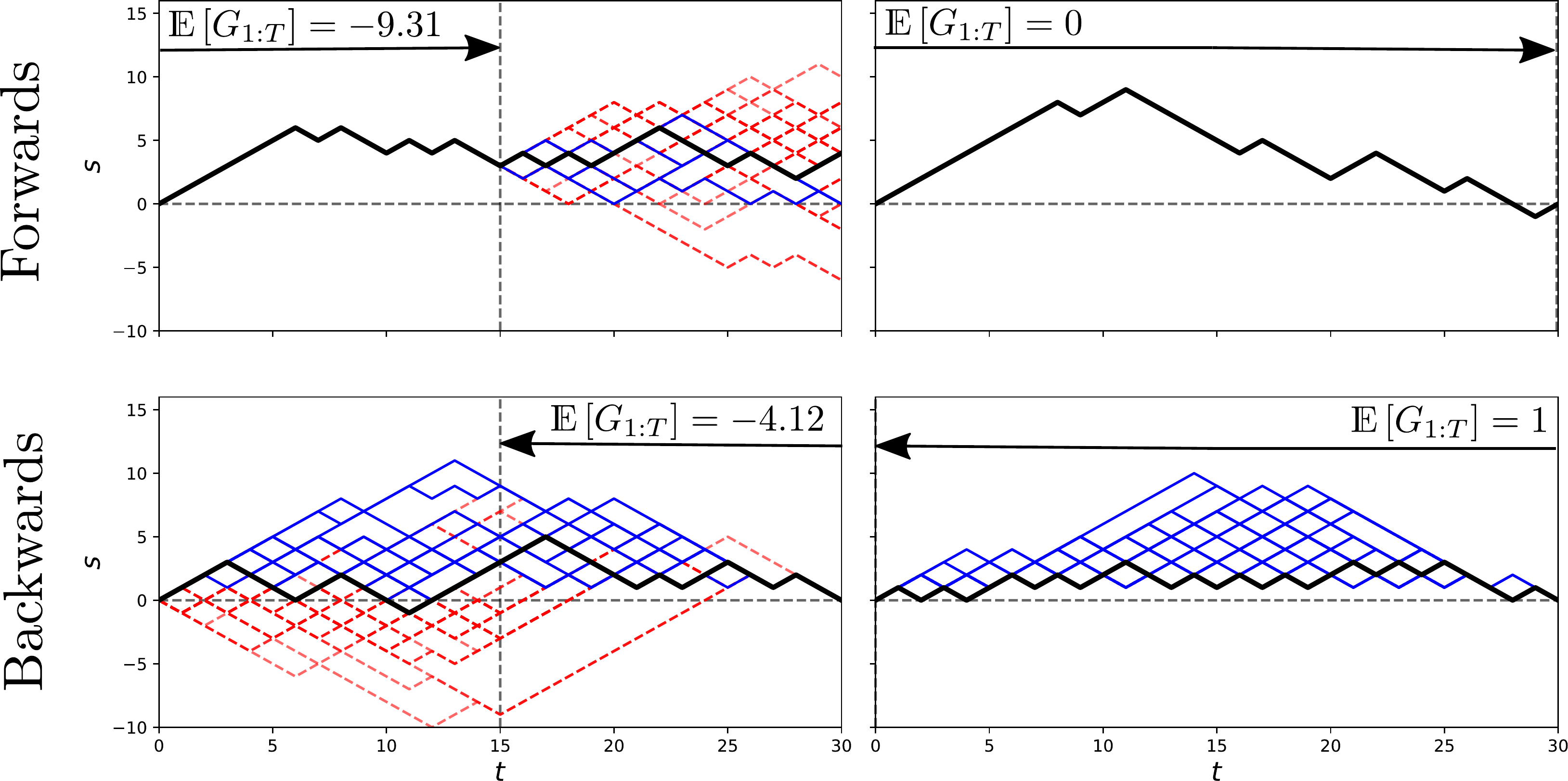}
    \caption{\textbf{Trajectories for the stochastic excursion problem generated by policies optimised using DMRG:} In each plot, $100$ trajectories are shown, generated by taking a random action (red or blue lines), in addition to a ''greedy" trajectory generated by taking the most probable action (thick black line) at each time-step, according to a policy optimised using DMRG. Solid blue lines satisfy the excursion conditions, while dashed red do not. In the first row policies are optimising using a forwards-in-time DMRG sweep, while in the second a backwards-in-time sweep is used. For each step in the DMRG algorithm, the constrained optimisation is achieved by parametrising the policy using $|S| \times |A|$ real numbers, which are scaled and normalised appropriately, and applying a gradient free optimisation method (Powell's method). The number of tensors that have been optimised to produce the policies are $T/2$ and $T$ for the columns respectively, as indicated by the vertical dashed lines and arrows. The expected return of the policy is also shown in each panel.}\label{fig:trajectories}
\end{figure*}

\section{Example: Conditioned Dynamics for Rare Trajectory Generation}
\label{sect:Conditioned_Dyn}

\subsection{Conditioned Dynamics and FMDPs}

Often when studying stochastic dynamics, the particular trajectories of interest occur only rarely
\cite{Bolhuis2002,Garrahan2018}. In many cases, analytical study of the trajectory statistics is intractable, and we must resort to numerical sampling. An important problem is finding an alternative dynamics which generates rare trajectories efficiently \cite{Borkar2003,Majumdar2015,Chetrite2015,Jack2019}. This can be phrased as an FMDP with an appropriate reward structure, such that the trajectories generated by its optimal policies satisfy the desired conditions on the original dynamics.

An elementary example of rare events are ``stochastic excursions'' \cite{Majumdar2015}, where a simple random walker is conditioned to stay above a certain line and at a given time must return to this line.
For a symmetric random walker, the probability of an excursion scales as $T^{-3/2}$. In terms of an FMDP, the conditioned dynamics can be encoded as the solution of an optimisation problem where movement below the zero line, or failure to reach it at $T$, is given a negative reward. Such a solution will be highly degenerate, and there are many different possible choices of reward structure that will lead to the same space of solutions.

For an episode with fixed termination time, $T$, the positions of the random walker are encoded in $\Ss = \lbrace -T,\ldots,-1,0,1,+T \rbrace$, such that $\dSs = 2 T + 1$. The action space is $\As = \lbrace 0, 1\rbrace$, where $a = 0, 1$ correspond to a down/up move of the walker, respectively. We assume the initial state distribution fixed at zero, $p(s_{0}) = \mathcal{I}\left[ s_{0} = 0\right]$, where $\mathcal{I}$ is the indicator function taking on the value one when the argument is true and zero otherwise. For illustration, we consider a dynamics where stochasticity is included only through the policy, though we emphasise that the TN method we apply for FDMPs has no such restriction. Indeed, one can consider not only general Markovian stochastic processes within the same framework, but also a variety of different rare events - such as meanders or bridges - by choosing the reward structure appropriately. 

For the generation of excursions, the reward structure we choose is as follows: when $t \neq T$, a reward of $0$ is given for $s \ge 0$ and $-1$ otherwise; at $t=T$, a reward of $1$ is given for $s=0$ and $-10$ otherwise. Thus, $\Rs = \lbrace 1, 0, -1, -10 \rbrace$ and the dynamics of the problem are governed by the following update rules: When $t \neq T$:
\begin{align}
s_{t} &= a_{t-1} (s_{t-1} + 1) + (1-a_{t-1}) (s_{t-1} - 1) ~ , \\
r_{t} &=  a_{t-1} \left( \mathcal{I}\left[ s_{t-1} \ge -1\right] -1 \right) + ~ , \nonumber \\
&~~ + \left(1-a_{t-1}\right) \left( \mathcal{I}\left[ s_{t-1} \ge 1\right] -1 \right) ~ ,
\end{align} 
and when $t = T$:
\begin{align}
s_{T} &= s^{+} ~ , \\
r_{T} &= a_{T-1}\left( 11 \, \mathcal{I}\left[ s_{T-1} = -1 \right] -10 \right) + \nonumber \\
& ~~~ + ( 1- a_{T-1}) \left( 11 \, \mathcal{I}\left[s_{T-1} = +1 \right] -10 \right) ~ .
\end{align}
By assumption, the system dynamics $p\left(s_{t},r_{t} | s_{t-1}, a_{t-1}\right)$ takes on the value one when the above relations are satisfied, and zero otherwise. Under a uniformly random policy - where up/down moves are equally likely regardless of the state - the dynamics will be that of an unconditioned random walker. Under an optimal policy, for which $\mathds{E}\left[G_{1:T}\right] = 1$, the walker will satisfy the excursion condition.

\subsection{DMRG for Excursions}

To illustrate the DMRG procedure for policy optimisation, we a consider a simple DMRG algorithm to solve the problem of generating excursions. Initially, a policy is generated by randomly selecting $T \times |\mathcal{S}| \times |\mathcal{A}|$ real numbers. These are scaled and normalised appropriately to form a valid policy, before proceeding with the policy optimisation. 

We perform a single sweep either forwards-in-time or backwards-in-time, thus optimising tensors $\pi_{1},\pi_{2},\ldots,\pi_{T}$ or $\pi_{T},\pi_{T-1},\ldots,\pi_{1}$, respectively. The total optimisation consists of $T$ iterations. At a given iteration, $n=1,\ldots,T$, the environment is evaluated by contracting all tensors in the network excluding $\pi_{n}$ (forward sweep) or $\pi_{T-n+1}$ (backward sweep). A constrained optimisation is then performed to minimise the negative expected return, shifted appropriately so that the minimum is zero. We apply a simple approach of gradient free optimisation (Powell's method), and satisfy the constraints directly by applying the necessary scaling and normalising to an input set of real values. While this method is slow, it is sufficient for this illustration, and can easily be replaced with more sophisticated ones for solving the necessary constrained optimisation.

Using the policy found by these optimisations, trajectories can be generated from the FMDP which obey the excursion condition, see Fig. \ref{fig:trajectories}. Trajectories sampled with four different polices are shown. In the first (second) row, the policy is determined via a forwards-in-time (backwards-in-time) sweep. How optimisation progresses is shown by the columns: in the first one, half the policy tensors have been optimised, and in the second one a full sweep has been completed. As can be seen, in the full-sweep backwards case (lower right)  all trajectories generated by the policy are excursions as expected (solid blue lines); in contrast, a single full-sweep forwards can fail to find a policy that generates excursions (top right) though randomly restarting the policy optimisation allows for one to post-select a deterministic policy that generates an excursion.

Additionally, the backwards sweep discovers a policy that is stochastic (non-deterministic), while the policy found during the forwards sweep is found to be deterministic in every case: a side effect of the degeneracy of the optimal policies.
Since in the backwards sweep the policy in the future of each step in the iteration is optimal, there are multiple actions which are seen to produce the same expected return, which the optimization algorithm does not uniquely focus on.
In contrast, due to the random initialization each step of the forward sweep sees a distinct expected return for each action it can take, and thus the optimization algorithm focuses precisely on whichever is currently seen as best according to the incorrect future policy.

\section{Conclusions}
\label{sect:conclusions}

We have introduced a tensor network formulation for Markov decision processes, along with an policy optimisation algorithm based on those usually applied to matrix product states. TNs and the associated optimisation algorithms are extremely flexible and can certainly be adapted accommodate more sophisticated cases beyond the class of MDPs considered here. Possible generalisations include: (i) termination time $T$ that can vary between episodes; (ii) continuing MDPs using uniform MPS/transfer matrix methods \cite{Vanderstraeten2019}; (iii) non-Markovian system dynamics or non-local in time reward structure, optimising for a non-Markovian policy; (iv) integration of TNs into standard RL algorithms, such as model-based approaches for unknown system dynamics, see e.g.\ \cite{Wang2019}, or using the a TNR as a natural model of the value function. As such, the formalism we present here lays the ground to pursue a number of avenues of research combining tensor networks with reinforcement learning more broadly.

\section*{Acknowledgement}
We thank N. Pancotti for useful discussion and reading of the manuscript. The  research leading  to  these  results  has  received  funding  from the  Leverhulme Trust  [grant  number  RPG-2018-181] and University of Nottingham grant no. FiF1/3. We are grateful for access to the University of Nottingham's Augusta HPC service. We acknowledge the use of Athena at HPC Midlands+.

\bibliography{tensor_network_approach_to_RL}
\bibliographystyle{icml2020}

\end{document}